\documentclass[nofootinbib]{revtex4}
\usepackage{graphicx}
\usepackage{amsmath}
\usepackage{amssymb}
\usepackage{float}
\usepackage{color}

\begin{document}
\title{Static general relativistic solutions supported by phantom and ordinary scalar fields with higher-order potentials}

\author{Vladimir Dzhunushaliev}
\email{v.dzhunushaliev@gmail.com}
\affiliation{
	Institute of Experimental and Theoretical Physics,  Al-Farabi Kazakh National University, Almaty 050040, Kazakhstan
}
\affiliation{
	Department of Theoretical and Nuclear Physics,  Al-Farabi Kazakh National University, Almaty 050040, Kazakhstan
}
\affiliation{
	Institute of Physicotechnical Problems and Material Science of the NAS of the Kyrgyz Republic, 265 a, Chui Street, Bishkek 720071,  Kyrgyzstan
}

\author{Vladimir Folomeev}
\email{vfolomeev@mail.ru}
\affiliation{
	Institute of Experimental and Theoretical Physics,  Al-Farabi Kazakh National University, Almaty 050040, Kazakhstan
}
\affiliation{
	Institute of Physicotechnical Problems and Material Science of the NAS of the Kyrgyz Republic, 265 a, Chui Street, Bishkek 720071,  Kyrgyzstan
}
\author{Arislan Makhmudov}
\email{arslan.biz@gmail.com}
\affiliation{Peoples Friendship University of Russia, Educational and Scientific Institute of Gravitation and Cosmology}

\author{
	Ainur Urazalina
}
\email{Y.A.A.707@mail.ru}
\affiliation{
	Department of Physics, K. Zhubanov Aktobe Regional State University, Aktobe, 030000, Kazakhstan
}


\begin{abstract}
Domain wall, wormhole, particlelike, and cosmic string general relativistic solutions supported
by two interacting phantom or ordinary scalar fields with 4th-, 6th-, and 8th-order potentials are studied.
Numerical calculations indicate that regular finite energy solutions exist only for specific values of two free parameters of the potentials.
By solving nonlinear eigenvalue problems for some fixed sets of values of the free parameters and of boundary conditions,
it is shown that the presence or absence of the solutions depends on a particular symmetry of the problem, on the type of the scalar fields
 (ordinary or phantom), and on the form of the potential.
\end{abstract}
\maketitle

\section{Introduction}

In recent years interest in obtaining solutions with various scalar fields
has grown considerably, primarily because of the discovery of the accelerated expansion of the present Universe.
It is now widely believed that such acceleration is caused by the presence of a special form of matter -- dark energy,
whose key feature is that it violates various energy conditions.

In the most extreme case, the violation of the so-called null energy condition can occur.
In hydrodynamical language, this corresponds to the fact that the effective pressure  of matter filling the Universe, $p$, is negative, and its modulus is
greater than the energy density $\varepsilon$, i.e. $p<-\varepsilon$. Such a substance is referred to as exotic matter.
As a model of exotic matter, one can consider phantom (or ghost) scalar fields, i.e., fields with the opposite sign in front of the kinetic term
of the scalar field Lagrangian density. Such fields are widely used both in describing the current accelerated expansion of the Universe~\cite{AmenTsu2010}
and in modelling various localized objects (see below).
The possible existence of phantom scalar fields in nature
is indirectly supported by the observed accelerated expansion
of the present Universe
(see, e.g., Refs.~\cite{Sullivan:2011kv,Ade:2015xua}, from which one may conclude that to explain the recent observational data
one should take exotic matter into consideration).

In the present paper we consider regular solutions to Einstein's gravitational equations supported by
two ordinary or phantom scalar fields.
In Ref.~\cite{Dzhunushaliev:2016}, we have obtained plane symmetric (domain walls), spherically symmetric (wormholes and phantom balls), and cylindrically symmetric (cosmic strings) solutions
supported by two interacting phantom scalar fields with a 4th-order potential. Here we extend those results and
study the possibility of obtaining such solutions with 6th- and 8th-order potential terms. Also, we compare the obtained results with those
found earlier for the 4th-order potential of Ref.~\cite{Dzhunushaliev:2016}.

The aforementioned configurations are well known in the bulk of literature.
Cosmic strings are extended objects that could be formed in the early Universe under phase transitions associated with spontaneous symmetry
breaking~\cite{Vilen,Bran2}. In their modelling various types of scalar fields are employed~\cite{Baze,Sant}, including two interacting scalar fields~\cite{BezerradeMello:2003ei,Dzhunushaliev:2007cs}.

Another category of extended objects are plane symmetric domain walls, which are topological defects that arise in both particle physics
and cosmology~\cite{Bran2,Cvet}. In particular,  domain wall solutions may exist in theories where a scalar field potential has isolated minima,
and a domain wall is a surface that separates those minima~\cite{thin_dom}. In such a case a scalar field changes over space and
tends asymptotically to two different minima.
The region where the scalar field changes rapidly corresponds to the domain wall.
In the thin-wall approximation, the change in the scalar field energy density is localized on the surface of the domain wall, and it is replaced
by a delta function~\cite{thick_dom}. In the case where all fields are constant on each side of the wall, i.e., when they are at the potential minimum, the domain walls are called vacuum domain walls.

Finally, one can consider a situation where scalar fields are localized
on  relatively small scales comparable to the sizes of stars.
In this case, they may create spherically symmetric configurations possessing both trivial and nontrivial spacetime topologies.
As an example of systems with a trivial topology, one can consider boson stars consisting of various ordinary scalar fields~\cite{Schunck:2003kk,Liebling:2012fv}.
In turn, the use of phantom scalar fields permits obtaining solutions of the Einstein-matter equations
describing configurations both with a trivial~\cite{Dzhunushaliev:2008bq} and
a nontrivial wormholelike topology (for a recent review on the subject, see, e.g., Ref.~\cite{Bronnikov:2018vbs}),
including configurations supported by complex ghost scalar fields \cite{Dzhunushaliev:2017syc}.

In this paper, we consider all four types of configurations (domain walls, particlelike systems, wormholes, and cosmic strings) constructed from two interacting phantom or ordinary
scalar fields with higher-order potentials. Systems with two ordinary scalar fields are well known from quantum field theory~\cite{rajaraman}.
In the presence of a gravitational field, such systems were also repeatedly considered in the cosmological and astrophysical
contexts~\cite{2_fields_syst}. In our previous papers we have obtained a number of solutions with two scalar fields (both ordinary and phantom ones)
which can be used both in describing astrophysical objects and when considering cosmological problems: regular spherically and cylindrically
symmetric solutions~\cite{2_fields_our,Dzhunushaliev:2007cs,Dzhunushaliev:2015sla}; cosmological solutions~\cite{Dzhunushaliev:2006xh,Folomeev:2007uw};
thick brane solutions supported by ordinary and phantom scalar fields~\cite{2_fields_brane}. In the present paper we proceed with research in this direction by considering scalar fields with different potentials.

\section{General equations}
\label{GE}

We consider compact gravitating configurations consisting of two real scalar fields $\phi$ and $\chi$.
The modeling is carried out  within the framework of Einstein's general relativity.
The corresponding Lagrangian of the system is (hereafter, we work in units where $c=\hbar=1$)
\begin{equation}
\label{lagrangian}
L=-\frac{R}{16\pi G}+\epsilon\left[
			\frac{1}{2}\partial_\mu \phi \partial^\mu
			\phi + \frac{1}{2}\partial_\mu \chi \partial^\mu \chi - V(\phi,\chi)
		\right],
\end{equation}
where $R$ is the scalar curvature, $G$ -- the Newtonian gravitational constant, $\mu, \nu=0,1,2,3$, and
 $\epsilon=+1$ or $-1$ corresponds to ordinary or phantom  fields, respectively.
Using this Lagrangian, the  gravitational and scalar field equations can be written in the form:
\begin{eqnarray}
\label{EinstEQ}
  R^k _i-\frac{1}{2}\delta^k_i R &=& \kappa T_i^k ,
\\
	\frac{1}{\sqrt{-g}}\frac{\partial}{\partial  x^i} \left[
	\sqrt{-g} g^{i k} \frac{\partial (\phi,\chi)}{\partial x^k}
	\right] &=& - \frac{\partial V}{\partial (\phi,\chi)} ~,
\label{scalar_eqs}
\end{eqnarray}
where $\kappa = 8\pi G$. In the present paper we assume that
the interacting scalar fields have a potential in one of the forms:
\begin{eqnarray}
\label{pot1}
	V(\phi,\chi) &=& \frac{\lambda_1}{4}(\phi^2 - m_1^2)^2 + \frac{\lambda_2}{4}(\chi^2 - m_2^2)^2 + \phi^2
	\chi^2 - V_0,
\\
\label{pot2}
	V(\phi,\chi) &=& \frac{\lambda_1}{2}\phi^2(\phi^2 - m_1^2)^2 + \frac{\lambda_2}{2}\chi^4(\chi^2 - m_2^2)^2+\frac{1}{2}\phi^2
	\chi^2 - V_0,
\\
\label{pot3}
	V(\phi,\chi) &=& \frac{\lambda_1}{4}\phi^4(\phi^2 - m_1^2)^2 + \frac{\lambda_2}{2}\chi^4(\chi^2 - m_2^2)^2 + \frac{1}{2}\phi^2
	\chi^2 - V_0.
\end{eqnarray}
Here,  $m_1$ and $m_2$ are some free parameters, $\lambda_1 $ and $\lambda_2 $ -- self-interaction constants, and $V_0$ -- a constant whose value can be chosen from the statement of the problem.

The corresponding energy-momentum tensor entering the right-hand side of Eq.~\eqref{EinstEQ} is
\begin{equation}
\label{emt}
	T_{\mu}^\nu = \epsilon \left\{
		\partial_\mu \phi \partial^\nu \phi+
		\partial_\mu \chi \partial^\nu \chi-
		\delta_{\mu}^\nu \left[
			\frac{1}{2}\partial_\rho \phi \partial^\rho
			\phi + \frac{1}{2}\partial_\rho \chi \partial^\rho \chi - V(\phi,\chi)
		\right]
	\right\}.
\end{equation}

\section{Domain walls }
\label{DW}

\begin{figure}[t]
			\begin{center}
							\includegraphics[width=1\linewidth]{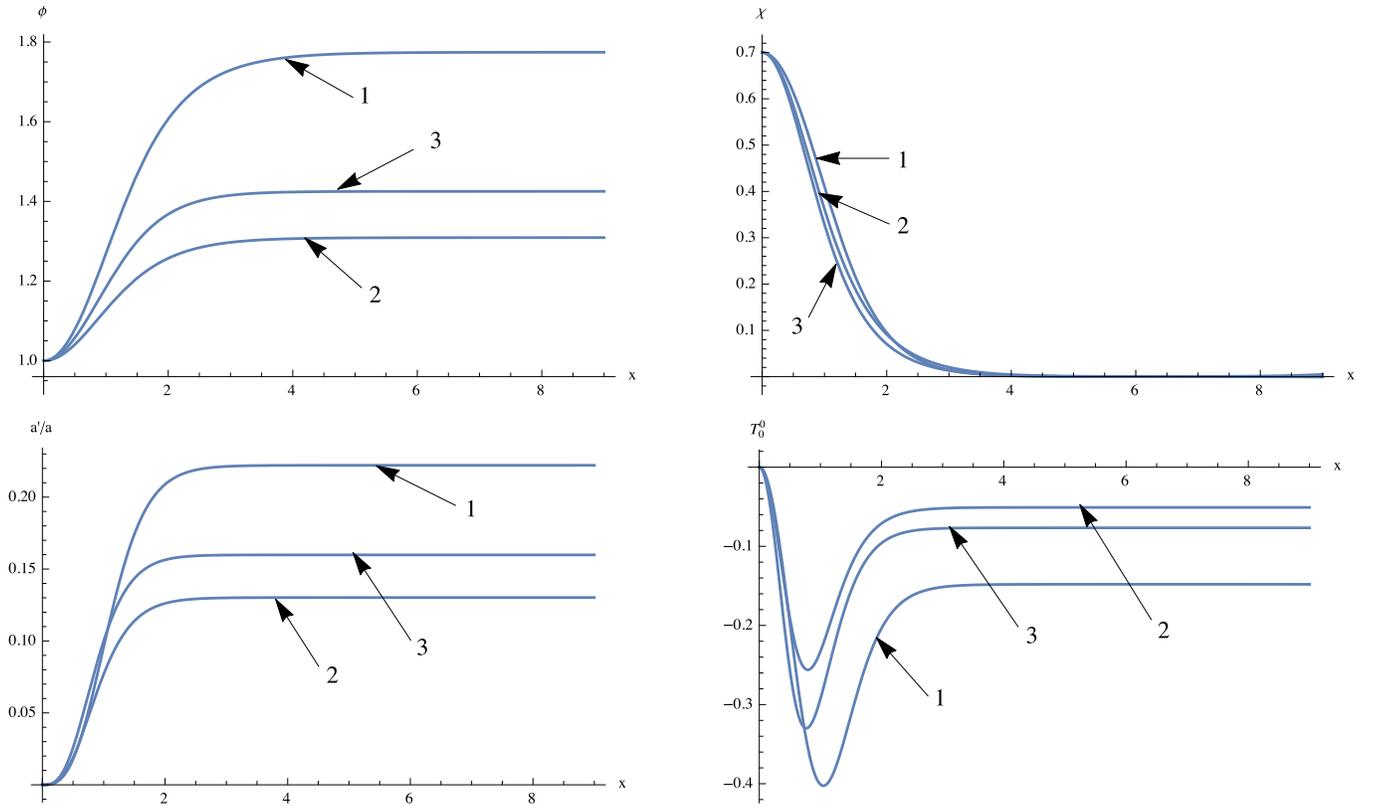}
			\end{center}
		\caption{Phantom domain wall:
profiles of the phantom ($\epsilon=-1$) scalar fields $\phi(x), \chi(x)$, the metric function $a^\prime(x)/a(x)$, and the energy density $T^0_0(x)$ are shown.
The curve 1 corresponds to the 4th-order potential~\eqref{pot1},
the curve 2 -- to the 6th-order potential~\eqref{pot2}, the curve 3 -- to the 8th-order potential~\eqref{pot3}. 
The labeling of the curves is also valid for all other figures presented below.}
		\label{fig_dom_wall_phantom}
\end{figure}

\begin{figure}[t]
			\begin{center}
							\includegraphics[width=1\linewidth]{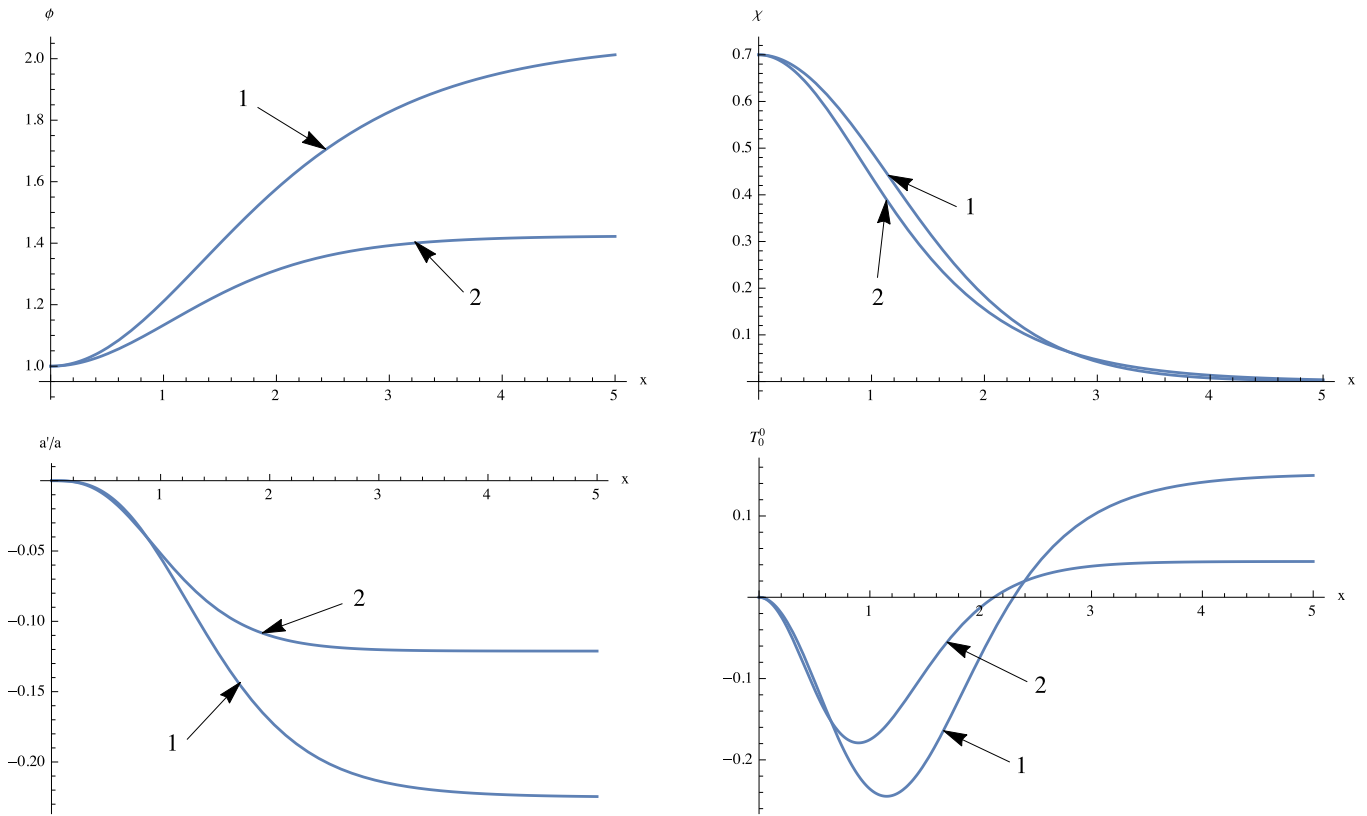}
			\end{center}
		\caption{Ordinary domain wall:
profiles of the ordinary ($\epsilon=+1$) scalar fields $\phi(x), \chi(x)$, the metric function $a^\prime(x)/a(x)$, and the energy density $T^0_0(x)$ are shown.
}
		\label{fig_dom_wall_ord}
\end{figure}

In considering plane symmetric domain walls solutions, we choose the metric in the form:
\begin{equation}
\label{metric_wall}
ds^2=a^2(x) (dt^2-dy^2-dz^2)-dx^2,
\end{equation}
where $x,y,z$ are Cartesian coordinates. Then Eqs.~\eqref{EinstEQ}-\eqref{emt} yield
\begin{eqnarray}
\label{ein_wall_1}
	3\left(\frac{a^\prime}{a}\right)^2 &=& -\epsilon\left[-\frac{1}{2}\left(
	\phi^{\prime 2} + \chi^{\prime 2}\right) + V\right] ~,
\\
	\frac{a^{\prime \prime}}{a} - \left(\frac{a^\prime}{a}\right)^2 &=& -\frac{\epsilon}{2}\left(\phi^{\prime 2} + \chi^{\prime 2}\right) ~,
\label{ein_wall_2} \\
\label{field_wall_1}
	\phi^{\prime \prime} + 3 \frac{a^\prime}{a}\phi^\prime &=& \phi\left[2\chi^2 + \lambda_1(\phi^2 - m_1^2)\right] ~,
\\
	\chi^{\prime \prime} + 3 \frac{a^\prime}{a}\chi^\prime &=&
	\chi\left[2\phi^2 + \lambda_2(\chi^2-m_2^2)\right] ~,
\label{field_wall_2}
\end{eqnarray}
where Eqs.~\eqref{ein_wall_1} and \eqref{ein_wall_2} are the $\left(_1^1\right)$ and $\left[\left(_0^0\right)-\left(_1^1\right)\right]$ components of
the Einstein equations, respectively, and the  prime denotes differentiation with respect to $x$.
The results of numerical calculations for the phantom ($\epsilon=-1$)  and ordinary $(\epsilon=+1)$  scalar fields
are shown in Figs.~\ref{fig_dom_wall_phantom} and~\ref{fig_dom_wall_ord}, respectively.
The corresponding eigenvalues of the parameters $m_{1,2}$ for the potentials \eqref{pot1}-\eqref{pot3} and $\epsilon = \pm 1$ are given in Table~\ref{eignvlsDW}.

\begin{table}[H]
	\begin{center}
		\begin{tabular}{ |c|c|c|c|c|}
			\hline
			\# &  Potentials                   & $\epsilon$                   & $m_1$                   & $m_2$
			\\
			\hline
			1   & 4th-order    & -1     & 1.77426601     & 1.80400455
			\\
			 \hline
			2   & 6th-order     & -1     & 1.30901092     & 1.73766048
			\\
			\hline
			3   & 8th-order      & -1    & 1.4251234264     & 1.7965336329
			\\
			\hline
			4   & 4th-order      & +1     & 2.05880064139     & 1.720175382122
			\\
			\hline
			5   & 6th-order      & +1     & 1.42405708294     & 1.61615084819
			\\
			\hline
			6   & 8th-order      & +1     & no solution     & no solution
			\\
			\hline
		\end{tabular}
	\end{center}
	\caption{Eigenvalues of the parameters $m_1, m_2$ for the phantom/ordinary domain wall solutions with the 4th-, 6th-, and 8th-order potentials~\eqref{pot1}-\eqref{pot3}.
The boundary conditions at the center $x=0$ are $\phi_0=1, \chi_0=0.7, a_0=1, \phi^\prime_0 =\chi^\prime_0= a^\prime_0=0$.
The values of the free parameters $\lambda_1=0.15, \lambda_2=1.1$.}
\label{eignvlsDW}
\end{table}

\section{Phantom balls}

\begin{figure}[t]
			\begin{center}
							\includegraphics[width=1\linewidth]{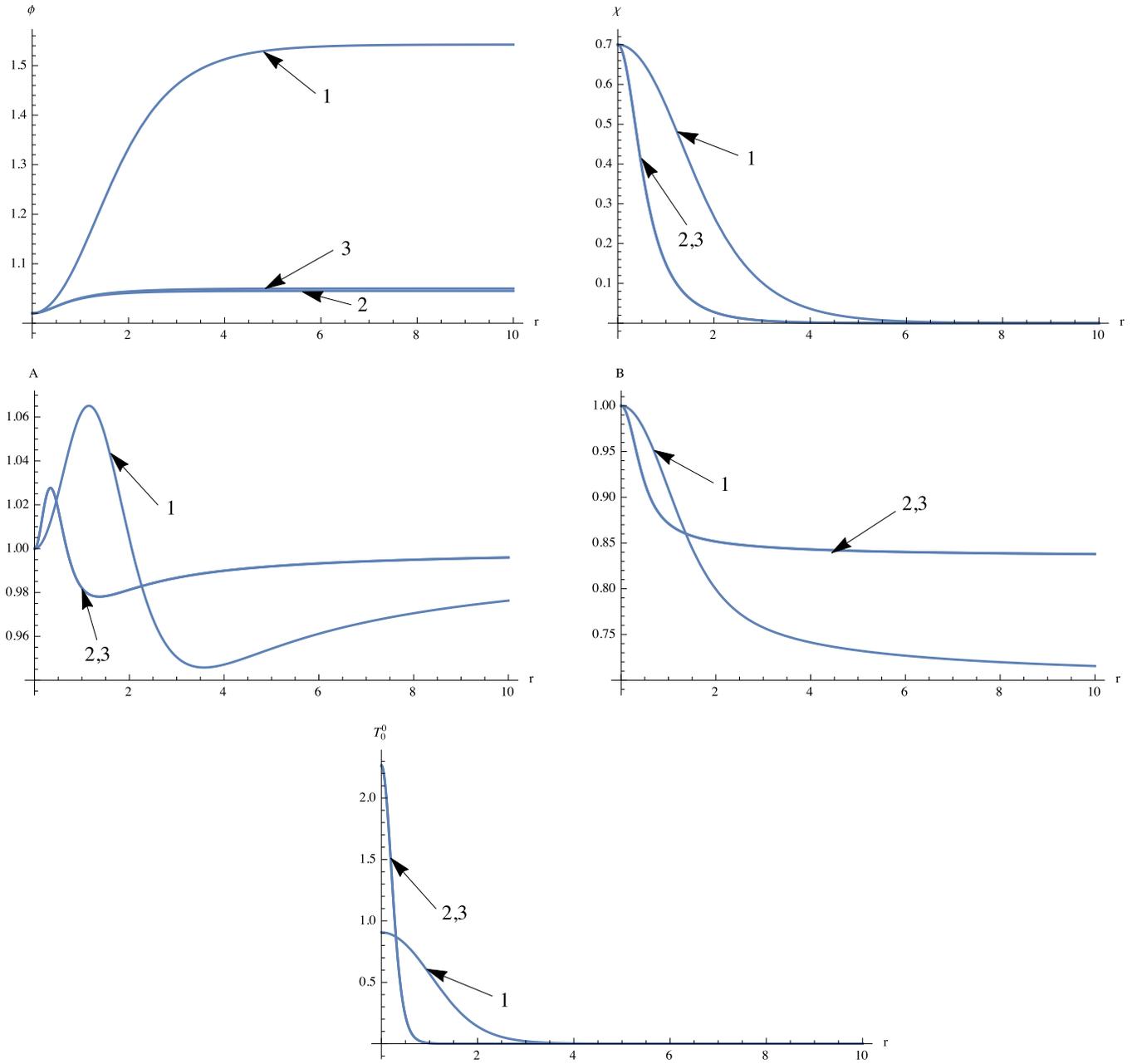}
			\end{center}
		\caption{ Phantom ball:
profiles of the scalar fields $\phi(r), \chi(r)$, the metric functions $A(r), B(r)$, and the energy density $T^0_0(r)$ are shown.
}
		\label{fig_phant_ball}
\end{figure}

Let us now consider particlelike solutions supported by phantom fields.
For this case,
we choose the spherically symmetric line element in Schwarzschild coordinates
\begin{equation}
\label{metric_sph}
ds^2 = B(r)dt^2 - A(r)dr^2 - r^2(d\theta^2 + \sin^2\theta d\varphi^2),
\end{equation}
where $r,\theta,\varphi$ are spherical coordinates. The Einstein and scalar field equations \eqref{EinstEQ} and \eqref{scalar_eqs}
 together with Eq.~\eqref{emt} give for the phantom case
\begin{eqnarray}
\label{einst1_sph}
	\frac{1}{r}\frac{A^\prime}{A^2} + \frac{1}{r^2}\left(1 -
	\frac{1}{A}\right) &=&
	 - \frac{1}{2A}\left(\phi^{\prime 2} + \chi^{\prime 2}\right) -
	V(\phi,\chi),
\\
	\frac{1}{r}\frac{B^\prime}{A B} -
	\frac{1}{r^2}\left(1 - \frac{1}{A}\right) &=&
	- \frac{1}{2A}\left(\phi^{\prime 2} +\
	\chi^{\prime 2}\right)+V(\phi,\chi),
\label{einst2_sph} \\
\frac{B^{\prime \prime}}{B} -
	\frac{1}{2}\left(\frac{B^\prime}{B}\right)^2 - \frac{1}{2}\frac{A^\prime}{A}\frac{B^\prime}{B}	- \frac{1}{r}\left(\frac{A^\prime}{A} -
	\frac{B^\prime}{B}\right) &=&  2A\left[\frac{1}{2A}\left(\phi^{\prime 2} + \chi^{\prime 2}\right) + V(\phi,\chi)\right],
\label{einst3_sph}
\end{eqnarray}
\begin{eqnarray}
\label{sfe1_sph}
	\phi^{\prime \prime} + \left(\frac{2}{r} + \frac{B^\prime}{2B} - \frac{A^\prime}{2A}\right)\phi^\prime &=&
	A\phi\left[2\chi^2+\lambda_1(\phi^2-m_1^2)\right]~,
\\
	\chi^{\prime \prime} + \left(\frac{2}{r} + \frac{B^\prime}{2B} - \frac{A^\prime}{2A}\right)\chi^\prime &=&
	A\chi\left[2\phi^2 + \lambda_2(\chi^2-m_2^2)\right]~,
\label{sfe2_sph}
\end{eqnarray}
where the  prime denotes differentiation with respect to $r$.
These equations describe spherically symmetric objects that can be called phantom balls \cite{Dzhunushaliev:2016}.
The results of numerical calculations are shown in Fig.~\ref{fig_phant_ball}.
Notice that the solutions for the 6th- and 8th-order potentials are practically coincide. Table~\ref{eignvls_bs} shows
the eigenvalues of the parameters $m_{1,2}$ for the potentials \eqref{pot1}-\eqref{pot3}.

\begin{table}[H]
	\begin{center}
		\begin{tabular}{ |c|c|c|c|}
			\hline
			\# &  Potentials                 & $m_1$                   & $m_2$
			\\
			\hline
			1   & 4th-order       & 1.54248223     & 1.89958804
			\\
			\hline
			2   & 6th-order        & 1.04506272     & 4.1962616
			\\
			\hline
			3   & 8th-order        & 1.050035     & 4.2023521
			\\
			\hline
		\end{tabular}
	\end{center}
	\caption{
Eigenvalues of the parameters $m_1, m_2$ for the phantom ball solutions with the 4th-, 6th-, and 8th-order potentials~\eqref{pot1}-\eqref{pot3}.
The boundary conditions at the center $r=0$
are $\phi_0=1, \chi_0=0.7, A_0=1, B_0=1, \phi^\prime_0 =\chi^\prime_0= B^\prime_0=0$.
The values of the free parameters $\lambda_1=0.15, \lambda_2=1.1$.
}
\label{eignvls_bs}
\end{table}

In the case of ordinary ($\epsilon=+1$) scalar fields and for the values of the parameters $\phi_0=1, \chi_0=0.7, \lambda_1=0.15, \lambda_2=1.1$,
we did not find solutions with the potentials~\eqref{pot1}-\eqref{pot3}.

\section{Wormhole solutions}

\begin{figure}[h]
			\begin{center}
							\includegraphics[width=1\linewidth]{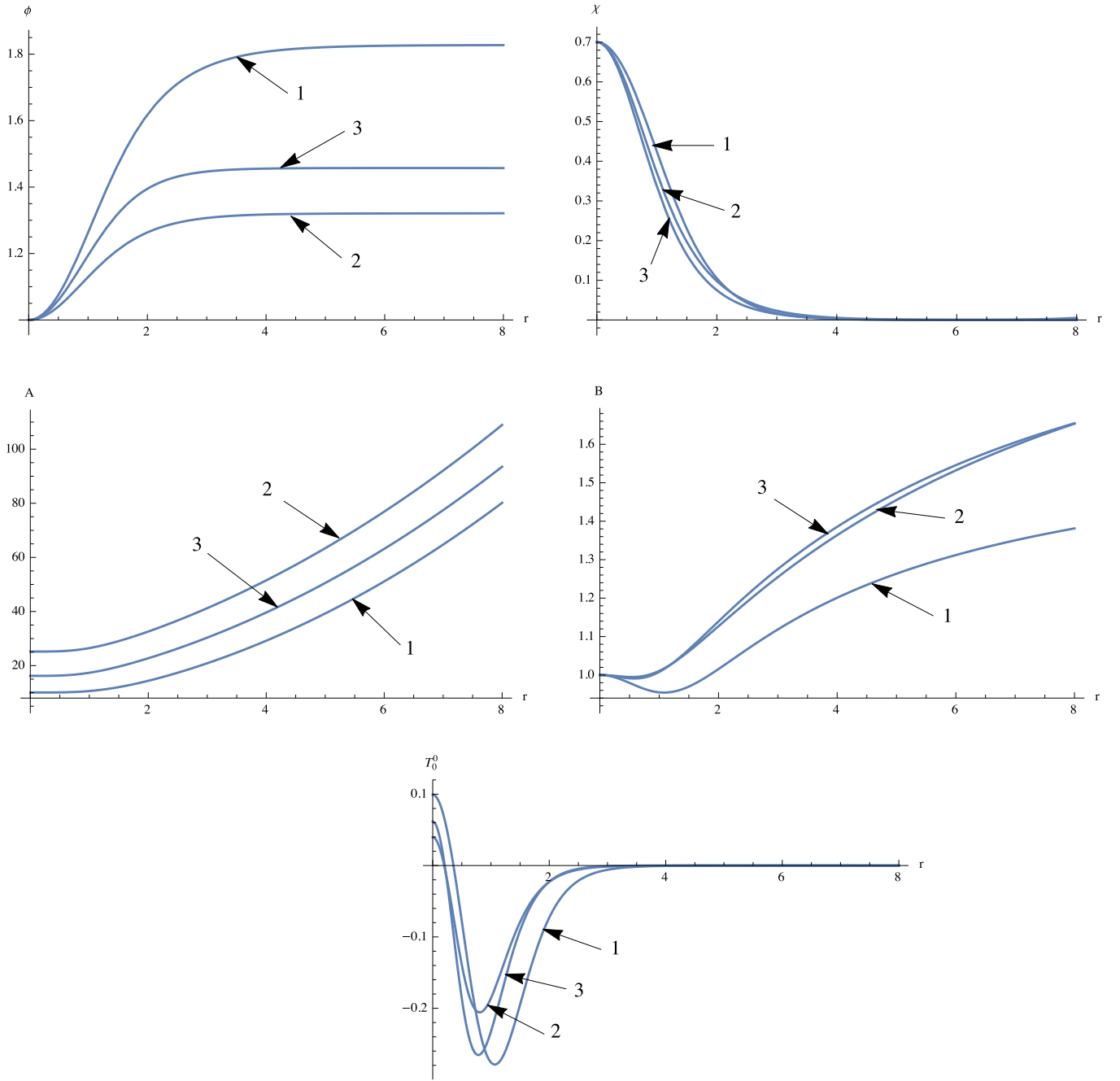}
			\end{center}
		\caption{Traversable wormhole:
profiles of the phantom scalar fields $\phi(r), \chi(r)$, the metric functions $A(r), B(r)$, and the energy density $T^0_0(r)$ are shown.
}
		\label{fig_wormhole}
\end{figure}

Here, it is convenient to choose the metric in polar Gaussian coordinates
\begin{equation}
\label{metric}
ds^2  = B(r) dt^2-dr^2-A(r)(d\theta^2+\sin^2\theta d\varphi^2),
\end{equation}
where $r,\theta,\varphi$ are spherical coordinates. Then one derives the following set of Einstein's and scalar field equations describing a traversable wormhole
supported by the phantom fields $\phi, \chi$:
\begin{eqnarray}
\label{Einstein_a}
	\frac{A^{\prime \prime}}{A} -
	\frac{1}{2}\left(\frac{A^{\prime}}{A}\right)^2 -
	\frac{1}{2}\frac{A^{\prime}}{A}\frac{B^{\prime}}{B} &=&
	\phi^{\prime 2} + \chi^{\prime 2}~,
\\
\label{Einstein_b}
	\frac{A^{\prime \prime}}{A} + \frac{1}{2}\frac{A^{\prime}}{A}\frac{B^{\prime}}{B} - \frac{1}{2}\left(\frac{A^{\prime}}{A}\right)^2  -
	\frac{1}{2}\left(\frac{B^{\prime}}{B}\right)^2 +
	\frac{B^{\prime \prime}}{B} &=& 2\left[\frac{1}{2}(\phi^{\prime 2} + \chi^{\prime 2}) + V\right] ,
\\
\label{Einstein_c}
	\frac{1}{4}\left(\frac{A^{\prime}}{A}\right)^2 - \frac{1}{A} + \frac{1}{2}\frac{A^{\prime}}{A}\frac{B^{\prime}}{B} &=&
	- \frac{1}{2}(\phi^{\prime 2} + \chi^{\prime 2}) + V ,
\\
\label{field_a}
	\phi^{\prime \prime} + \left(\frac{A^\prime}{A} + \frac{1}{2}\frac{B^\prime}{B}\right)\phi^\prime &=&
	\phi \left[2\chi^2 + \lambda_1(\phi^2 - m_1^2)\right] ,
\\
\label{field_b}
	\chi^{\prime \prime} + \left(\frac{A^\prime}{A} + \frac{1}{2}\frac{B^\prime}{B}\right)\chi^\prime &=&
	\chi \left[2\phi^2 + \lambda_2(\chi^2 - m_2^2)\right],
\end{eqnarray}
where the  prime denotes differentiation with respect to $r$.
The results of numerical calculations  are shown in Fig.~\ref{fig_wormhole}
for the eigenvalues of the parameters $m_{1,2}$ given in Table~\ref{eignvls_WH}.

\begin{table}[H]
	\begin{center}
		\begin{tabular}{ |c|c|c|c|}
			\hline
			\# &  Potentials                 & $m_1$                   & $m_2$
			\\
			\hline
			1   & 4th-order       & 1.82729811     & 1.7869422825
			\\
			\hline
			2   & 6th-order        & 1.32067169     & 1.7205753
			\\
			\hline
			3   & 8th-order        & 1.45731329     & 1.7806672
			\\
			\hline
		\end{tabular}
	\end{center}
	\caption{Eigenvalues of the parameters $m_1, m_2$ for the wormhole solutions with the 4th-, 6th-, and 8th-order potentials~\eqref{pot1}-\eqref{pot3}.
The boundary conditions at the throat $r=0$
are $\phi_0=1, \chi_0=0.7, A_0=-1/V\left(\phi_0,\chi_0\right), B_0=1, \phi^\prime_0 =\chi^\prime_0= A^\prime_0= B^\prime_0=0$.
The values of the free parameters $\lambda_1=0.15, \lambda_2=1.1$.
}
\label{eignvls_WH}
\end{table}

\section{Cosmic strings}
\label{DW}

\begin{figure}[t]
			\begin{center}
							\includegraphics[width=1\linewidth]{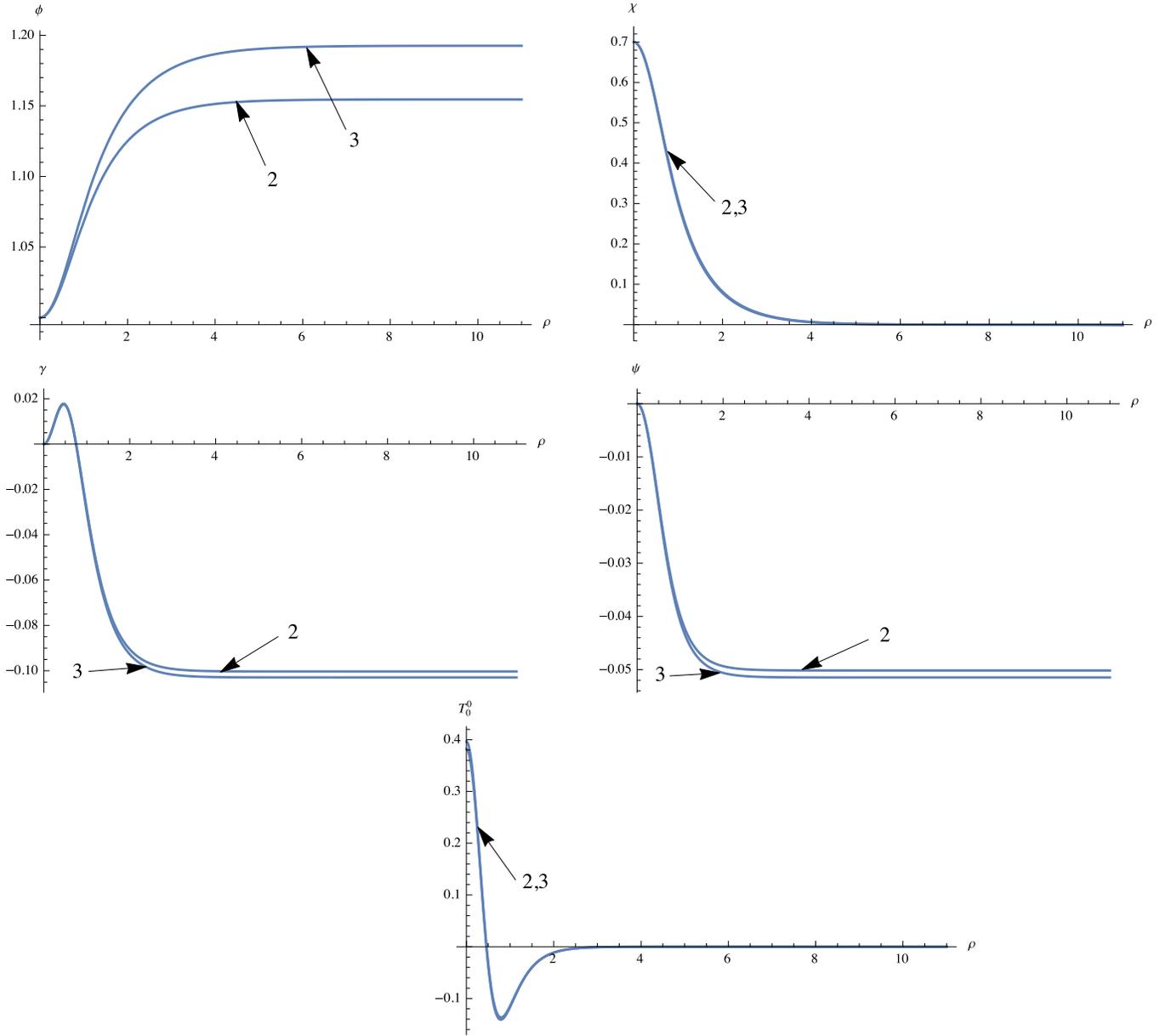}
			\end{center}
		\caption{Phantom cosmic string:
profiles of the scalar fields $\phi(\rho), \chi(\rho)$, the metric functions $\gamma(\rho), \psi(\rho)$, and the energy density $T^0_0(\rho)$ are shown.
}
		\label{fig_string}
\end{figure}

In describing such cylindrically symmetric objects, we use the metric
\begin{equation}
\label{metric_string}
ds^2 = e^{2 \nu(\rho)} dt^2 - e^{2 (\gamma(\rho) - \psi(\rho))} d\rho^2 -
e^{2 \psi(\rho)} dz^2 - \rho^2 e^{-2 \psi(\rho)} d\varphi^2 ,
\end{equation}
where $\rho,z,\varphi$ are cylindrical coordinates. In this case, one has the following Einstein and phantom scalar field equations:
\begin{eqnarray}
	\frac{\gamma^\prime}{\rho} - {\psi^\prime}^2 &=&
	- \kappa \left(
		\frac{1}{2} {\phi^\prime}^2 + \frac{1}{2} {\chi^\prime}^2 +
		e^{2 (\gamma - \psi)} V(\phi, \chi)
	\right) ,
\label{ef_1} \\
	\frac{\nu^\prime + \psi^\prime}{\rho} - {\psi^\prime}^2 &=&
	- \kappa \left(
		\frac{1}{2} {\phi^\prime}^2 + \frac{1}{2} {\chi^\prime}^2 -
		e^{2 (\gamma - \psi)} V(\phi, \chi)
	\right) ,
\label{ef_2} \\
	\psi^{\prime \prime } - \nu^{\prime \prime} - \psi^\prime \gamma^\prime +
	\nu^\prime \gamma^\prime - {\nu^\prime}^2 +
	\frac{\psi^\prime + \gamma^\prime - \nu^\prime}{\rho} &=&
	\kappa \left(
		- \frac{1}{2} {\phi^\prime}^2 - \frac{1}{2} {\chi^\prime}^2 -
		e^{2 (\gamma - \psi)} V(\phi, \chi)
	\right) ,
\label{ef_3} \\
	- \psi^{\prime \prime } - \nu^{\prime \prime} +
	\psi^\prime \gamma^\prime +
	\nu^\prime \gamma^\prime -
	2 {\psi^\prime}^2 - 2\psi^\prime \nu^\prime -
	{\nu^\prime}^2 &=& \kappa \left(
		- \frac{1}{2} {\phi^\prime}^2 - \frac{1}{2} {\chi^\prime}^2 -
		e^{2 (\gamma - \psi)} V(\phi, \chi)
	\right) ,
\label{ef_4}
\end{eqnarray}
\begin{eqnarray}
	\phi^{\prime \prime} + \phi^\prime \left(
		\frac{1}{\rho} - \gamma^\prime + \psi^\prime + \nu^\prime
	\right) &=& e^{2 (\gamma - \psi)} \phi \left[
		2 \chi^2 + \lambda_1 \left( \phi^2 - m_1^2 \right)
	\right] ,
\label{ef_5} \\
	\chi^{\prime \prime} + \chi^\prime \left(
	 \frac{1}{\rho} - \gamma^\prime + \psi^\prime + \nu^\prime
	\right) &=& e^{2 (\gamma - \psi)} \chi \left[
	 2 \phi^2 + \lambda_2 \left( \chi^2 - m_2^2 \right)
	\right] .
\label{ef_6}
\end{eqnarray}

To simplify them, let us make an additional assumption that
two of the metric functions are equal,  i.e., $\nu = \psi$. After some algebraic manipulations and performing the rescaling
$\rho/\sqrt \kappa \rightarrow \rho$, $\phi \sqrt \kappa \rightarrow \phi$, $\chi \sqrt \kappa \rightarrow \chi$,
and $m_{1,2} \sqrt \kappa \rightarrow m_{1,2} $, we get the following equations for the metric functions
$\gamma(\rho), \psi(\rho)$ and the phantom $(\epsilon=-1)$ scalar fields $\phi(\rho), \chi(\rho)$:
\begin{eqnarray}
\frac{\gamma^\prime}{\rho} - {\psi^\prime}^2 &=&
- \left(
\frac{1}{2} {\phi^\prime}^2 + \frac{1}{2} {\chi^\prime}^2 +
e^{2 (\gamma - \psi)} V(\phi, \chi)
\right) ,
\label{cosmic_string_1} \\
2 \frac{\psi^\prime}{\rho} - {\psi^\prime}^2 &=&
- \left(
\frac{1}{2} {\phi^\prime}^2 + \frac{1}{2} {\chi^\prime}^2 -
e^{2 (\gamma - \psi)} V(\phi, \chi)
\right) ,
\label{cosmic_string_1a} \\
\psi^{\prime \prime } + \frac{\psi^\prime}{\rho} &=&
e^{2 (\gamma - \psi)} \left(
1 - 2 \rho \psi^\prime
\right) V(\phi, \chi) ,
\label{cosmic_string_2} \\
\phi^{\prime \prime} + \phi^\prime \left(
\frac{1}{\rho} - \gamma^\prime + 2 \psi^\prime
\right) &=& e^{2 (\gamma - \psi)} \phi \left[
2 \chi^2 + \lambda_1 \left( \phi^2 - m_1^2 \right)
\right] ,
\label{cosmic_stringl_3} \\
\chi^{\prime \prime} + \chi^\prime \left(
\frac{1}{\rho} - \gamma^\prime + 2 \psi^\prime
\right) &=& e^{2 (\gamma - \psi)} \chi \left[
2 \phi^2 + \lambda_2 \left( \chi^2 - m_2^2 \right)
\right] ,
\label{cosmic_string_4}
\end{eqnarray}
where the prime denotes differentiation with respect to the rescaled radial coordinate $\rho$.

The results of numerical calculations are shown in Fig.~\ref{fig_string} for the eigenvalues of the parameters $m_{1,2}$ given in Table~\ref{eignvlsCS}.
It is seen that the profiles for $\chi$ and for the energy densities of the 6th- and 8th-order potential cases are practically coincide.

\begin{table}[H]
	\begin{center}
		\begin{tabular}{ |c|c|c|c|}
			\hline
			\# &  Potentials                 & $m_1$                   & $m_2$
			\\
			\hline
			1   & 4th-order       & no solution    & no solution
			\\
			\hline
			2   & 6th-order        & 1.154579476     & 2.30250731
			\\
			\hline
			3   & 8th-order        & 1.1926167892     & 2.32316842475
			\\
			\hline
		\end{tabular}
	\end{center}
	\caption{Eigenvalues of the parameters $m_1, m_2$ for the phantom cosmic string solutions with the 6th- and 8th-order potentials~\eqref{pot2} and \eqref{pot3}.
The boundary conditions at $\rho=0$
are $\phi_0=1, \chi_0=0.7, \psi_0=\gamma_0= \phi^\prime_0 =\chi^\prime_0= \psi^\prime_0=0$.
The values of the free parameters $\lambda_1=0.15, \lambda_2=1.1$.
}
	\label{eignvlsCS}
\end{table}

In the case of ordinary ($\epsilon=+1$) scalar fields and for the values of the parameters $\phi_0=1, \chi_0=0.7, \lambda_1=0.15, \lambda_2=1.1$,
we did not find solutions with the potentials \eqref{pot1}-\eqref{pot3}.



Summarizing the results,
we have obtained plane, cylindrically, and spherically (particlelike and wormhole) symmetric static general relativistic solutions supported by two interacting phantom/ordinary scalar fields
with 4th-, 6th-, and 8th-order potentials of the form~\eqref{pot1}-\eqref{pot3}. All the solutions have been obtained numerically for the fixed central values of the scalar fields
 $\phi_0=1, \chi_0=0.7$ and for the free parameters $\lambda_1=0.15, \lambda_2=1.1$.
 In doing so, we have solved nonlinear eigenvalue problems for the parameters $m_1, m_2$. It was shown that solutions may exist (or not exist) depending on a particular symmetry of the problem,
 on the type of the scalar fields (ordinary or phantom), and on the form of the potential.

\section*{Acknowledgments}
V.D. and V.F. gratefully acknowledge support provided by Grant No.~BR05236494
in Fundamental Research in Natural Sciences by the Ministry of Education and Science of the Republic of Kazakhstan. We are also grateful to the Research Group
Linkage Programme of the Alexander von Humboldt Foundation for the support of this research.

\end{document}